\documentstyle[12pt]{article}

\newcommand{\eq}{\begin{equation}}
\newcommand{\en}{\end{equation}}

\newcommand{\eqa}{\begin{eqnarray}}
\newcommand{\ena}{\end{eqnarray}}
\newcommand{\eqan}{\begin{eqnarray*}}
\newcommand{\enan}{\end{eqnarray*}}

\newcommand{\lbl}{\label}

\newcommand{\PL}[1]{Phys. Lett.\ {\bf #1}\ }

\newcommand{\PR}[1]{Phys. Rev\ {\bf #1}\ }

\def\sqr#1#2{{\vcenter{\hrule height.#2pt
     \hbox{\vrule width.#2pt height#1pt \kern#1pt
        \vrule width.#2pt}
     \hrule height.#2pt}}}

\def\thinspace{\kern .16667em}

\def\Dir{\nabla\kern-7.8pt\Big{/}}

\def\reali{{\hbox{l\kern-.5ex R}}}
\def\naturali{{\hbox{\s@ l\kern-.5mm N}}}
\def\interi{{\mathchoice
 {\hbox{ Z\kern-1.5mm Z}}
 {\hbox{ Z\kern-1.5mm Z}}
 {\hbox{{ Z\kern-1.2mm Z}}}
 {\hbox{{ Z\kern-1.2mm Z}}}  }}

\def\unity{{\hbox{\s@ 1\kern-.8mm l}}}
\def\uno{{\hbox{ 1\kern-.8mm l}}}
\def\part{\partial}

\newcommand{\modu}[1]{{\mid #1 \mid}}

\def\aa{\alpha}
\def\bb{\beta}
\def\cc{\chi}

\def\dd{\delta}
\def\DD{\Delta}

\def\etab{\bar\eta}
\def\ff{\phi}

\def\gg{\gamma}
\def\GG{\Gamma}

\def\ll{\lambda}
\def\LL{\Lambda}

\def\oo{\omega}

\def\pp{\psi}

\def\pb{\bar\psi}
\def\rr{\rho}

\def\sv{\vec{S}}
\def\fv{\vec{\ff}}
\def\G{{\cal G}}
\def\n0{{n\rightarrow0}}

\begin{document}
\begin{titlepage}
\begin{flushright}
NBI-HE-93-06\\
DFTT 1/93\\
April 1993\\
hep-lat/yymmnn
\end{flushright}
\vspace*{0.5cm}
\begin{center}
{\bf
\begin{Large}
{\bf
COLOURED POLYMERS\\}
\end{Large}
}
\vspace*{1.5cm}
         {\large Igor Pesando}\footnote{E-mail PESANDO@NBIVAX.NBI.DK,
22105::PESANDO, 31890::I\_PESANDO}
         \\[.5cm]
          The Niels Bohr Institute\\
          Blegdamsvej 17, DK-2100 Copenhagen \O \\
          Denmark
\footnote{and Dipartimento di Fisica Teorica dell' universit\`a di
Torino and
Istituto Nazionale di Fisica Nucleare, Sezione di Torino
         via P.Giuria 1, I-10125 Torino, Italy }
\end{center}
\vspace*{0.7cm}
\begin{abstract}
{
We show that non-oriented coloured polymers (self--avoiding walks
with different types of links) are in the same universality class of
the ordinary self--avoiding walks, while the oriented coloured are not.
}
\end{abstract}
\vfill
\end{titlepage}

\setcounter{footnote}{0}
In this letter we would consider the critical behaviour of coloured
polymers, both oriented and unoriented.
A coloured polymer is a polymer with different
kinds of monomers that have different binding energies among them;
these coloured polymers will be described, following De Gennes's
approach (\cite{DG}), by a self--avoiding walk with different colours on links.

We are interested in considering such models because:
\begin{itemize}

\item{1)} the unoriented coloured polymers model
is not obtainable with a generalization of De Gennes model
(\cite{DG}) but only generalizing the Hilhorst one
(\cite{Hi})\footnote{At least these two generalized models are not equivalent
off criticality as we will see.}. And
before the analytic continuation down to $n=0$, the Hilhorst
model and the generalized one  belong to different universality classes:
the model, we
present, has $n C$ components (hereafter $C$ indicate the number of colours)
with global symmetry $S_{n C}\otimes \interi_2$ \footnote{
$S_n$ is the symmetric group of $n$ elements.} while
Hilhorst model has $n$ components with $S_n \otimes \interi_2$ symmetry.
Hence it is not, a priori, completely clear whether the analytic
continuation to $n=0$ will bring the two model in the same universality
class as one could expect naively.
\item{2)} as we will explain below, the model we consider gives meaning
to negative statistical weights.
\item{3)} they seem  to be more realistic that De Gennes' model
(\cite{DG}), since in
nature there are polymers that are not composed by one kind of monomer
only.
\item{4)} the coloured oriented polymers model can be in a
different class of universality  with respect to the usual polymers.
\end{itemize}

The model, we consider for unoriented polymers,
is defined by the following partition function:
\eq
Z_n=Tr \exp(K \sum_{i,j\in\G} \sum_{a,b=1}^C \sv_{a i}\cdot A_{i j}\sv_{b j}
{}~w_{ab})
\lbl{Z}
\en
where $K$ is the temperature of the model,
$A_{i j}$ is the adjacency matrix of the lattice (or more generally the
graph ) $\G$
\footnote{We use $i,j,k\dots$ to indicate a site of the lattice $\G$,
$\aa, \bb, \gg \dots$ to denote the components of the vector $\sv$},
$\sv_{a i}\equiv (S_{a i}^\aa)$ is a $n$-components vector defined on the
site $i$ with colour $a$,
$(w)_{a b}=w_{ab}$ is the matrix of statistical weights that determines the
probability of the transition from colour $a$ to $b$; we point out that
$w_{a b}=w_{b a}$.
The trace is defined as follows
\eq
Tr(\bullet)=\int~\prod_{i\in\G} ~\prod_{a=1}^C
d\sv_{a i}~{1\over n C}\sum_{\aa,a}
\dd(S_{a i}^{\aa~2}-n C)\prod_{\bb\ne\aa, a}\dd(S^\bb_{a i})~\bullet
\lbl{traccia}
\en
This expression means that only one component of the vector $(\sv_{1
i}\dots\sv_{C i})\equiv\oplus_{a=1}^{a=C}\sv_{a i}$ is different from
zero ant its value can be $\pm \sqrt{n C}$; hence
the average on a site $i$ is done on $2n C$ configurations
characterized geometrically by the fact that they are the centers of the
faces of a hypercube of side $2\sqrt{n C}$ in $n C$ dimensions.

Following Hilhorst we get immediately that in the limit $\n0$
\eqa
<S^\aa_{a i} S^\bb_{b i}>&=&\dd^{\aa\bb}\dd_{a b}
\nonumber\\
<S^{\aa_1}_{a_1 i} \dots S^{\aa_n}_{a_n i}>&=&0~~\forall n>2
\ena
It is also easy to find that
\eq
\lim_{\n0} Z_n =1
\lbl{z0}
\en

\noindent
and that the two points function yields at $n=0$
\eq
<S^\aa_{a i} S^\bb_{b j}>=
\dd^{\aa\bb}\sum_L N_{a i;b j}(L)~K^L
\lbl{gen-fun}
\en
where $N_{a i;b j}(L)$ is the number of  self--avoiding walks
that have ends in $i$ and $j$ where there are colours $a$ and $b$ respectively;
explicitly we have
\eq
N_{a i;b j}(L)=N_{i; j}(L) ~ (w^L)_{a b}
\lbl{N(L)}
\en

Would we have used the De Gennes approach, i.e. if we would have used as
density $\rr(S_{ i})\propto\prod_a\dd(\sv_{a i}^2-n)$ in the definition
of the trace,
we still had found (\ref{z0}) but we had not found (\ref{gen-fun})
because, differently from the one colour De Gennes model, here not only
the self--avoiding walks would contribute to (\ref{gen-fun}) but also
$C$ tolerant walks (i.e. walks that can self intersect at most $C$
times): this happens because
we have different colours, and, for instance, the trace
$Tr_i(S^\aa_{i a}S^\aa_{i a}S^\bb_{i b}S^\bb_{i b})\propto (1-\dd_{a b})$
is different from zero using De Gennes approach
and it allows for configurations like:

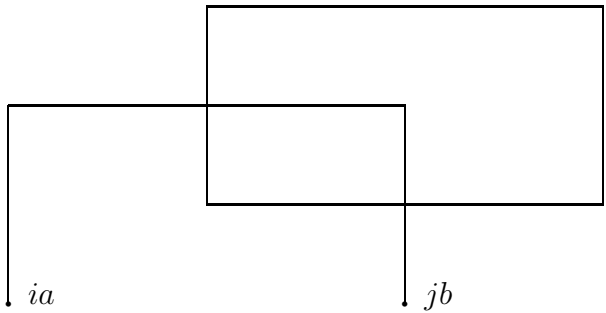
\setlength{\unitlength}{0.75pt}
\begin{picture}(300,200)(-50,-25)
\put(0,0){\circle*{2}}
\put(10,0){$i a$}
\put(200,0){\circle*{2}}
\put(210,0){$j b$}

\put(200,0){\line(0,1){100}}
\put(200,100){\line(-1,0){200}}
\put(0,100){\line(0,-1){100}}

\put(100,50){\line(1,0){200}}
\put(300,50){\line(0,1){100}}
\put(300,150){\line(-1,0){200}}
\put(100,150){\line(0,-1){100}}
\end{picture}

\noindent
nevertheless the critical behaviour of $C$ tolerant coloured walks
is the same of the coloured self avoiding walks exactly as it happens
to usual SAW and k--tolerant walks(\cite{Sh}).

In order to prove this affirmation,
we notice that any symmetric real $N\times N$ matrix can be
diagonalized by an $SO(N)$ rotation, in particular we can choose a
$R\in SO(C)$ such that
\eq
w=R^T ~\LL~ R
\en
where $\LL=diag(\ll_1\dots\ll_C)$ is the matrix of eigenvalues of $w$,
that can be either positive or null or negative:
with this meaning this model gives an interpretation to
negative weights also ( see also eq. (\ref{Z-diagonale}) ).
Redefining
\eq
\sv_{i a}'=R\sv_{i a}
\lbl{redefinizione}
\en
we can immediately rewrite (\ref{Z})
as
\eq
Z_n=Tr \exp(K \sum_{i,j\in\G} \sum_{a=1}^C \ll_a \sv_{a i}'\cdot A_{i j}
\sv_{b j}')
\lbl{Z-diagonale}
\en
Before we can perform a gaussian transform on this expression we have to
take care of the antiferromagnetic couplings using the standard way:
firstly we divide the lattice $\cal G$ (supposed not frustrated) in two
sublattices $\cal G^{(+)}$ and $\cal G^{(-)}$ in such a way each site of
the first (second) sublattice can be reached by an even number of steps
from the sublattice itself and by an odd number from the second (first)
sublattice, then we flip the sign of all the spins in the first
sublattice getting in this way a ferromagnetic coupling\footnote{
The difference between AF and F coupling can be found in the fact that
for AF coupling the magnetic field is not uniform but has opposite
values on the two sublattices $\cal G^{(+)}$ and $\cal G^{(-)}$}.
If we perform the gaussian transform we get
\eq
Z_n=cst~\int \prod_{i\in\G}\prod_{a=1}^{C}d\fv_{a i}~
e^{-{1\over 4K}\sum_{i,j\in\G}\sum_a \fv_{a i} \cdot A^{-1}_{i j}\fv_{a j}
     +\sum_i V_{eff}(\sqrt{\modu{\ll}}\fv_i)}
\en
where in the Hilhorst case the effective potential is
\eqa
e^{-V_{eff}|_i}\strut_{H}
&=&1+\sum_{m=1}^{\infty}{(n C)^{m-1}\over (2 m)!}
     \sum_{a,\aa} (\sqrt{\modu{\ll_a}}\ff_{a i}^\aa)^{2m}
\nonumber\\
&=&1
+{1\over2}\sum_{a,\aa} \modu{\ll_a}(\ff_{a i}^\aa)^{2}
+{n C\over24}\sum_{a,\aa} \ll_a^2(\ff_{a i}^\aa)^{4}
+\dots
\ena
while in the De Gennes case it turns out to be
\eq
e^{-V_{eff}|_i}\strut_{DG}
=\prod_a\left(1
+{1\over2}\sum_\aa \modu{\ll_a} (\ff_{a i}^\aa)^{2}
+{n\over 8(n+2)} \ll^2_a\left(\sum_\aa(\ff_{a i}^\aa)^{2}\right)^2
+\dots
\right)
\en
At the critical dimension we can consider only the relevant operators and
the effective actions become\footnote{Notice the coupling constants related to
the null eigenvalues of $\LL$ are obviously zero.}
\eqa
S_{eff~H}&=&\int~d^D x
      ~\sum_a \left[\fv_a\cdot(-\DD+m^2_a)\fv_a
                    +u_a (\fv^2_a)^2
                     +n~v_a\sum_\aa\ff_{\aa a}^4 \right]
\nonumber\\
S_{eff~DG}&=&\int~d^D x
       ~\sum_a \left[\fv_a\cdot(-\DD+m^2_a)\fv_a +g_a (\fv^2_a)^2
\right]\ena
that in the limit $\n0$ yield the same result because the cubic anisotropy
coefficient is proportional to $n$.
Moreover at the critical point only the fields with zero mass
propagate: these have the maximum absolute value of the eigenvalues
$\ll_M=\max\{\modu{\ll_a}\}$ since $m^2_a\propto {1\over 4Kq}
-{1\over 2}\modu{\ll_a}$.
We will denote by
${\cal F}=\{a|1\le a\le C\mbox{ s.t. }\modu{\ll_a}=\ll_M\}$
the set of the indices of the critical fields and with ${\cal F}^\pm$
the subsets
${\cal F}^\pm=\{a|1\le a\le C\mbox{ s.t. }\ll_a=\pm\ll_M\}$.

Let us now reconsider the asymptotic behaviour of the 2-point functions
(\ref{gen-fun}) that can be now rewritten as
\eqa
<S^\aa_{a i} S^\bb_{b j}>&=&
\sum_p R_{p a} R_{p b}<S^\aa_{p i} ~'~ S^\bb_{p j}~'>
\sim
\sum_{p\in{\cal F}} R_{p a} R_{p b}<\ff^\aa_{p i} ~'~ \ff^\bb_{p j}~'>
\nonumber\\
&\sim&{\dd^{\aa\bb}
      \left(
        \sum_{p\in{\cal F}^+} R_{p a} R_{p b}
         +(-)^{\modu{i-j}}\sum_{p\in{\cal F}^-} R_{p a} R_{p b}
      \right)
\over r_{i j}^{D-2+\eta}}
\lbl{gen-fun-diagonale}
\ena
where we used the well known fact (\cite{AdVS})
that both F and AF share the same critical exponent $\eta$.
{}From what follows we immediately deduce that the coloured
polymers have the same critical exponents of usual SAW.
There is another way of getting the same result and it consists on
considering the generating function of all the coloured polymers having
a fixed end:
\eq
\chi^{\aa\bb}_{a b}=
\sum_j <S^\aa_{a i} S^\bb_{b j}>=
\sum_p R_{p a} R_{p b}\sum_j <S^\aa_{p i} ~'~~ S^\bb_{p j}~'>=
\sum_p R_{p a} R_{p b}~\dd^{\aa\bb}\sum_L N(L)~(K~\ll_p)^L
\en
the first singularity happens when $K\ll_M={1\over z}$, where $z$ is defined
by the asymptotic behaviour $N(L)\sim_{L>>1} z^L L^{\gg-1}$, and hence
we get
\eq
\chi^{\aa\bb}_{a b}=
\sum_j <S^\aa_{a i} S^\bb_{b j}>
\sim \dd^{\aa\bb}
     \sum_L
      \left(
        \sum_{p\in{\cal F}^+} R_{p a} R_{p b}
         +(-)^L\sum_{p\in{\cal F}^-} R_{p a} R_{p b}
      \right)
     (z K\ll_M)^L L^{\gg-1}
\en

We now show that if the matrix $w$ is symmetric and not block diagonal then
all the fields become critical to the same "temperature", i.e.
$\forall a,b~~\sum_{p\in{\cal F}} R_{p a} R_{p b}\ne0$.
This is easily demonstrated if we note that a symmetric matrix $w$ that is not
block diagonal is irreducible\footnote{
In the subsequent text a vector with positive (non-negative) coordinates
will be called a positive (non-negative) vector, and similarly for a
matrix.

A matrix $A$ is called reducible if there is a permutation matrix $P$
such that $P^{-1}A P$ is of the form
$  \left(\begin{array}{cc} X & Y \cr
                          0 & Z  \end{array}\right)
$, where $X$ and $Z$ are square matrices. Otherwise, $A$ is called
irreducible.}, the vectors $\vec{r}_p=(R_{p a})$ at fixed $p$
are eigenvectors of $w$  and we apply the Perron-Frobenius theorem\footnote{
The part of the Perron-Frobenius theorem, we need, states that:
An irreducible non--negative matrix $A$ always has a positive
eigenvalue $r$ that is a simple root of the characteristic polynomial.

The modulus of any other eigenvalue does not exceed $r$.

To the "maximal" eigenvalue $r$ there corresponds a positive
eigenvector.

Moreover, if $A$ has $h$ eigenvalues of modulus $r$, then these numbers
are all distinct and are roots of the equation $\ll^h-r^h=0$.
}for the non-negative matrices (\cite{Ga}): in particular it follows
that the sets ${\cal F}^{\pm}$ are made just of one element.

One could still wonder whether a model for oriented self--avoiding
walks, that could admit an asymmetric weight matrix $w$, could exhibit
also different critical behaviour: the answer is yes.
An asymmetric weight matrix can be generated if we allow for different
binding energies for each of the two (different) ends of a monomer:
for instance, the weight matrix for a one-colour oriented model
can be written as
\eq
w=
  \left(\begin{array}{cc} w_{++} & w_{+-} \cr
                          w_{-+} & w_{++}  \end{array}\right)
\lbl{2x2}
\en
where $\pm$ distinguish between the two possible orientations of the
monomer within the oriented polymer, and we used the fact that
$w_{--}=w_{++}$ as it is shown by the following picture where
$e_1,e_2$ are the two different endpoints of the monomer

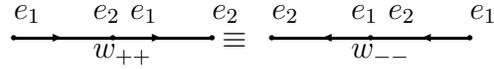
\begin{picture}(300,50)(-50,-25)

\put(20,00){\circle*{2}}
\put(20,10){$e_1$}
\put(20,00){\line(1,0){50}}
\put(45,00){\vector(1,0){0}}

\put(70,00){\circle*{2}}
\put(60,10){$e_2~e_1$}
\put(60,-10){$w_{++}$}

\put(70,00){\line(1,0){50}}
\put(95,00){\vector(1,0){0}}
\put(120,00){\circle*{2}}
\put(120,10){$e_2$}

\put(125,-5){$\equiv$}

\put(150,00){\circle*{2}}
\put(150,10){$e_2$}
\put(150,00){\line(1,0){50}}
\put(175,00){\vector(-1,0){0}}

\put(200,00){\circle*{2}}
\put(190,10){$e_1~e_2$}
\put(190,-10){$w_{--}$}

\put(200,00){\line(1,0){50}}
\put(225,00){\vector(-1,0){0}}
\put(250,00){\circle*{2}}
\put(250,10){$e_1$}

\end{picture}

The model for oriented coloured SAW is  defined by the partition
function (a generalization of (\cite{P}) obtained through
the approach of (\cite{Sh1}))
\eq
Z_{Oriented~Coloured~SAW}=
Tr \exp\left(K\sum_{i,j\in\G}\sum_{a=1}^C
   (a_{a i} A_{i j} P_{a j}+
    \etab_{a i} A_{i j} \eta_{a j})
       \right)
\lbl{Z-topo-0}
\en
where $a_{a i}$, $P_{a i}$ are commuting fields and
$\eta_{a i}$, $\etab_{a i}$ are grassman fields.
The non vanishing traces are, by definition:
\eqa
Tr 1&=&1
\nonumber\\
Tr a_{a} P_{b}&=& Tr \eta_a\etab_b= w_{a b}
\ena
Using these definitions it is easy to show that
\eqa
Z_{Oriented~Coloured~SAW}=1
\nonumber\\
<a_{a i} P_{b j}>=
\sum_L N_{a i;b j(oriented)}(L)~K^L
\ena
where $N_{a i;b j(oriented)}(L)$ is analogous to the formula
(\ref{N(L)}).
If we perform a gaussian transformation on (\ref{Z-topo-0}),we find
$$
Z_{Oriented~Coloured~SAW}=
\int~\prod_{a=1}^C~\prod_{i\in\G}
{}~d\ff_{a i}~d\oo_{a i}~d\pp_{a i}~d\pb_{a i}
\left( 1-i K w_{a b}(\oo_{a i}\ff_{b i}+ \pb_{a i}\pp_{b i}) \right)
$$
\eq
e^{i~\sum_{i,j}\sum_a A_{i j}^{-1}(\oo_{a i}\ff_{a j}+ \pb_{a i}\pp_{a j}) }
\lbl{Z-topo}
\en
where  $\oo_{aj}$ and $\ff_{aj}$ are bosonic fields defined in $j\in \G$
with "oriented" colour $a$ and similarly for the grassmann fields
$\pb_{a j}$ and $\pp_{a j}$.
It can be easily shown that near the criticality
\eq
w_{ac}<\ff_{c i}\oo_{d j}>w_{d b}\propto
\sum_L N_{a i;b j(oriented)}(L)~K^L
\en
In the model defined by (\ref{Z-topo}) we can use both asymmetric and
symmetric $w$, but differently from the symmetric case where
by means of a field redefinition we can always put the action in a diagonal
form, in the general asymmetric case we cannot diagonalize the quadratic
part of the action: this happens because an asymmetric real matrix can
 at most put in the Jordan canonical form, i.e.
$w=R
  \left(\begin{array}{ccc} J_1 & 0 & 0  \cr
                            0  & \dots & 0 \cr
                            0  & \dots &  J_p
 \end{array}\right)R^{-1}
$
where the generic Jordan block is of the form
$J_r=
  \left(\begin{array}{cccc} \ll_r & 1 & 0 & \dots \cr
                            0 & \ll_r & 1 & \dots \cr
                            0 & 0 & \ll_r & \dots \cr
                              & \dots & \dots &
 \end{array}\right)
$.
If we perform the naive continuum limit of (\ref{Z-topo}) and we throw
away the most massive fields, we get
\eq
S_{eff}=\int~d^D x~
i\sum_{a,b}\left(\oo_{a }(-\DD\dd_{a b}+M_{a b})\ff_{b }+
            \pb_{a }(-\DD\dd_{a b}+M_{a b})\pp_{b}\right)
-\left(\sum_{a,b} G_{a b} (\oo_{a }\ff_{b }+ \pb_{a }\pp_{b }) \right)^2
\en
where $M$ and $G$ are two upper triangulated invertible
matrices with all the diagonal element equal \footnote{Notice that
this action is stable under the symmetries
$$\oo\rightarrow \rr\oo~~,~~\ff\rightarrow
 {1\over\rr}\ff~~,~~\rr\in\reali$$
and
$$\oo\rightarrow U\oo~~,~~\ff\rightarrow V\ff$$
where both $U$ and $V$ are two upper triangulated invertible matrices
with all the diagonal element equal}.

Since a positive matrix $w$ , equivalent to a Jordan block, has all the
eigenvalues equal, it follows from the Perron-Frobenius theorem that the
matrix must be reducible, more precisely applying recursively the
Perron-Frobenius theorem to the square blocks of the equivalent
triangular matrix, it is easy to show that the most general form for
such a $w$ is
\eq
w
=P^T U P
=P^T  \left(\begin{array}{cccc} \ll & u_{1 2} & u_{1 3}  & \dots \cr
                            0 & \ll & u_{2 3} & \dots \cr
                            0 & 0 & \ll & \dots \cr
                            0  & \dots & \dots & \ll
 \end{array}\right) P
\lbl{w-gen}
\en
For a Jordan matrix $J$, we also find that
\eq
(J^n)_{ab}=\left\{\begin{array}{lcl}
\left(^{\,\,\,n}_{b-a}\right)\ll^{b-a}&\mbox{if}& a\le b
\mbox{\,\,and\,\,} b-a<n\cr
0 &\mbox{\,}& \mbox{otherwise}
                  \end{array}\right.
\lbl{J-ab}
\en
and this implies that the susceptibility $\cc_{ab}$ can be written as
(supposing for simplicity that the weight matrix can be put in
a Jordan canonical form consisting of just one Jordan block)
\eq
\cc_{a b}=
  R_{a c} \sum_L N(L) J^L_{c d}~R^{-1}_{d b}
\sim
  \sum_{d\ge c} R_{a c} R^{-1}_{d b} {\GG(\gg+d-c)\over(d-c)!~\ll^{d-c}}
  {1\over \left({1\over\log(z\ll K)}\right)^{\gg+d-c} }
\en
where we used the asymptotic form of $N(L)$, the Stirling formula for
the factorial, the formula (\ref{J-ab}) and
$\sum_L x^L L^{\aa-1}
\sim{\GG(\aa) \over \left(\log{1\over x}\right)^\aa}$.
\def\DB{{\bar\DD}}
Now if we call $\bar \DD$ the biggest $\DD=d-c$ for which
$\sum_{c=1}^{c=C-\DD}R_{a c}R^{-1}_{c+\DD,b}\ne0$, we find the critical
behaviour
\eq
\cc_{a b}
\sim
  {\GG(\gg+\DB)\sum_{c=1}^{c=C-\DB} R_{a c} R^{-1}_{c+\DB,b} \over\DB!~\ll^\DB}
  {1\over (-\log(z\ll K))^{\gg+\DB}}
\en
{}From the formula (\ref{w-gen}) and from $w^n=P^{-1}U^n P$, it follows
that the $\cc_{a a}$ have the usual critical exponent while the possible
unusual critical exponents can be find in $\cc_{a b}$ with $a\ne b$.
We can understand intuitively this phenomenon.
Consider the oriented one-colour case (\ref{2x2}), where $w_{-+}=0$,
then all the polymers beginning and ending with the same colour ($\pm$)
are just made of monomers of the same colour as the ends.
On the contrary a polymer beginning with $+$ and ending by $-$ is made gluing
a $+$ polymer and a $-$ polymer, in such a way that the sum of the
length of the two polymers is equal to the fixed total length, and this
can be done in as many ways as it is the total length.

In conclusion the critical behaviour of
the non oriented coloured polymers is completely equivalent to
the behaviour of the usual polymers, while the behaviour of the oriented
(coloured) polymers can be different, even if there is just one set of
critical exponents from which the others can be derived.

\end{document}